\documentclass[twocolumn,showpacs,preprintnumbers,amsmath,amssymb]{revtex4}
\usepackage{graphicx}% Include figure files
\usepackage{dcolumn}% Align table columns on decimal point
\usepackage{bm}% bold math
\usepackage{epsfig}
\usepackage{rotate}
\usepackage{lscape}

\begin{document}

\title{Coulomb screening and collective excitations in biased bilayer graphene}
\author{Xue-Feng Wang}
\email{xf_wang1969@yahoo.com}
\affiliation{Department of physics,
Soochow University, 1 Shizi Street, Suzhou, China 215006}
\author{Tapash Chakraborty}
\affiliation{Department of Physics and Astronomy, The University
of Manitoba, Winnipeg, Canada, R3T 2N2}

\begin{abstract}
We have investigated the Coulomb screening properties and plasmon spectrum in a bilayer
graphene under a perpendicular electric bias. The bias voltage applied between the two
graphene layers opens a gap in the single particle energy spectrum and modifies the
many-body correlations and collective excitations. The energy gap can soften the
plasmon modes and lead to a crossover of the plasmons from a Landau damped mode to being
undamped. Plasmon modes of long lifetime may be observable in experiments and may have
potentials for device applications.
\end{abstract}
\maketitle

Bilayer graphene (BLG) has attracted much attention due to its unique electronic characteristics,
distinct from the Dirac gas in monolayer graphene and the Fermi gas in traditional semiconductor
quantum wells \cite{mcca,wang2,hwan}. In addition, an energy gap between the conduction and valence
bands of a BLG can be opened and tuned by introducing an electrostatic potential bias between the
two graphene layers \cite{oost,zhan,stau,min,gava,falk,mcca1,mcca2}. This can be easily realized
via one or more external gates to perpendicularly bias BLG and make it a potential component for
integrated electronics. It is then very intriguing to understand some fundamental properties
such as correlation and screening properties of electron gases in a biased BLG. As collective
excitations, plasmon modes are a direct result of electronic correlation due to Coulomb interaction
between electrons. Experimental detection of plasmon modes has recently become feasible and has
been used to determine the dynamical behavior of electrons
in graphene layers \cite{liu,kram,bost1}.

Previously, assuming zero or non-zero spin-orbit interaction induced energy gap, we have studied
the Coulomb screening and collective excitation spectrum of intrinsic and doped monolayer graphenes
at zero and finite temperatures in the random phase approximation (RPA) \cite{wang}. Later, Qaiumzadeh
and Asgari \cite{qaiu} assumed an unspecified energy gap of arbitrary width for doped monolayer
graphene and studied the corresponding ground-state properties at zero temperature in RPA.
They concluded that the conductance and charge compressibility decrease with the band gap. Furthermore,
a THz source has been proposed based on the stimulated plasmon emission in graphene \cite{rana} and
the absorption of THz electromagnetic radiation in gapped graphene has been estimated \cite{wrig}. On
the other hand, the Coulomb screening and the collective excitations in zero gap BLG have been studied
in our previous work at zero and finite temperatures \cite{wang2} and by Hwang and Das Sarma \cite{hwan}
for the zero temperature case. In this paper we report on our studies of the correlations, screening,
and the plamson spectrum of electron gases in a biased BLG.

In the effective-mass approximation \cite{mcca}, the Hamiltonian describing electrons of moderate
energies in the $K$ valley of a biased BLG reads
\begin{equation}
H_K=\frac{\hbar^2}{2m^\ast} \left(
\begin{array}{cc}
0& k_-^2\\
k_+^2 & 0\\
\end{array}
\right) +\frac{U}{2} \left(
\begin{array}{cc}
1& 0\\
0 & -1\\
\end{array}
\right)
\end{equation}
with $k_\pm=k_x\pm i k_y=-i\nabla_x\mp i\nabla_y$ and ${\bm k}=(k_x,k_y)$ being measured from the
$K$ point. The effective mass of the quadratic term is $m^{\ast}=2\hbar^2\gamma_1/(3a_0\gamma_0)^2
\approx 0.033 m_0$ with $m_0$ the free electron mass, $a_0=1.42$ \AA\ the C-C bond length on the
graphene layer, $\gamma_0= 3.16$ eV the intra-layer coupling, and $\gamma_1=0.4$ eV the direct
inter-layer coupling. The second term arises from the electrostatic potential bias $U$ between the
two graphene layers separated by a distance $d=3.35$ \AA. The Hamiltonian is obtained by keeping
only the linear term of the Tayler expansion on the small energy value in unit of $\gamma_1$,
so it is valid for electrons of energy less than 0.4 eV which is adequate in our case. The indirect
inter-layer coupling is neglected since it affects only the energy band in the range of less than
2 meV from the middle of the conduction-valence band gap \cite{wang2}.

The eigenenergy of the above Hamiltonian is $E_{\bm{k}}^\lambda=\lambda U \sqrt{1+(\hbar^2k^2/m^\ast
U)^2}/2$ with the eigenfunctions $\Psi^{+1}_{\bm{k}}(\bm{r})={\small\left( \array{c}
\cos(\alpha_{\bm{k}}/2)\\
-\sin(\alpha_{\bm{k}}/2) e^{i2\theta_{\bm{k}}}
\endarray \right)}e^{i\bm{k}\cdot\bm{r}}$
and
$\Psi^{-1}_{\bm{k}}(\bm{r})={\small\left( \array{c} \sin(\alpha_{\bm{k}}/2)\\
\cos(\alpha_{\bm{k}}/2)e^{i2\theta_{\bm{k}}} \endarray \right)}e^{i\bm{k}\cdot\bm{r}}$
for $\lambda=+1$ and $-1$ respectively. Here $\theta$ is the azimuth of the vector $\bm{k}$,
i.e., $\tan\theta_{\bm{k}}=k_y/k_x$, and $\alpha_{\bm{k}}$ indicates the ratio
of the kinetic energy to the potential bias with $\tan\alpha_{\bm{k}}=\hbar^2k^2/(m^\ast U)$. The
conduction band which touches the valence band at $k=0$ in unbiased BLG becomes separated from
it by an energy gap equal to the potential bias $U$. This gap converts the BLG from a semimetal
into a semiconductor and accordingly modifies the optical and electric properties of the electrons
inside. For finite $U$, the density of states of the BLG diverges on the edge of the energy gap
$|E|=U/2$. At zero temperature, the carrier density $N$ in a BLG of Fermi energy $E_F$ is $N=\pm
\frac{2m^\ast}{\pi}\sqrt{E_F^2-U^2/4}$.

Following the well-established formalism for spin systems \cite{wang1}, we obtain the dielectric
matrix of a biased BLG in the form of a unit matrix multiplied by a dielectric function
$\varepsilon(q,\omega)=1-v_q\Pi ({\bf q},\omega) \label{dielectric}$ with the bare Coulomb interaction
$v_q=e^2/(2\varepsilon_0q)$ and the electron-hole propagator
\begin{equation}
\Pi({\bf q},\omega) =4\sum_{\lambda, \lambda', \bm{k}}
|g_{\bm{k}}^{\lambda,\lambda'}(\bm{q})|^2
\frac{f(E^{\lambda'}_{\bm{k}+\bm{q}})-f(E^\lambda_{\bm{k}})}
{\omega+E^{\lambda'}_{\bm{k}+\bm{q}}-E^\lambda_{\bm{k}}+i\delta}.
\label{propagator}
\end{equation}
%\end{widetext}
The factor four comes from the degenerate two spins and two valleys at $K$ and $K'$, $f(x)$ is the
Fermi function, and the vertex factor reads $|g^{\lambda,\lambda'}_{\bm{k}}(\bm{q})|^2=
\frac{1}{2}[1+\lambda\lambda'\cos\alpha_{\bm{k}}\cos\alpha_{\bm{k}+\bm{q}}
+\lambda\lambda'\sin\alpha_{\bm{k}}\sin\alpha_{\bm{k}+\bm{q}}
\cos(2\theta_{\bm{k}}-2\theta_{\bm{k}+\bm{q}})].$
At $q=0$ or $q=-2k$, $|g^{\lambda,\lambda'}_{\bm{k}}(\bm{q})|^2=(1+\lambda\lambda')/2$.
Similar to unbiased BLG, the interband vertical and back scatterings are both forbidden but
the intraband back scattering is allowed in biased BLG.

It has been shown that the interlayer indirect C-C interaction introduces anisotropic fine structures
near the Fermi energy in the range of 2 meV and leads to some interesting dielectric and collective
phenomena \cite{wang2}. For systems with energy gap $U> 5$ meV or with Fermi energy $E_F$ satisfying
$|E_F-kT|>3$ meV, this anisotropy becomes negligible. For large $U$ comparable to $\nu_1$, the effect
of the "Maxican hat" at the bottom (top) of the conduction (valence) band \cite{mcca} should be taken
into account. Nevertheless, for moderate $U$ and $E_F$, the model described here should be valid.
Furthermore, we assume that the BLG is far enough from the substrate and the gate so a unit
background dielectric constant is used in the calculation.

\begin{figure}
\vspace*{-0.5cm}
%\hspace*{-0.3cm}
\includegraphics*[height=130mm, width=90mm]{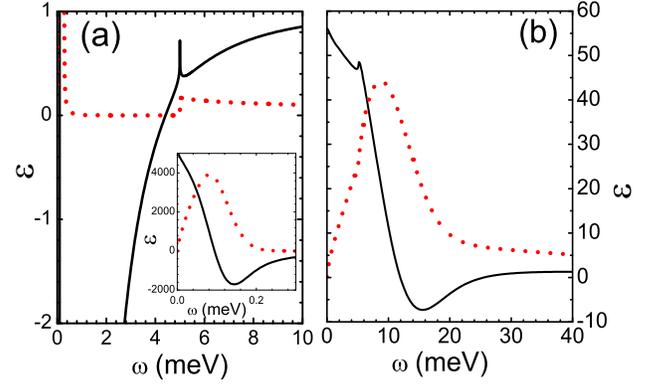}
\vspace{-8cm} \caption{$\varepsilon_r$ (solid) and $\varepsilon_i$ (dotted) are plotted versus $\omega$
in biased intrinsic BLG at temperature T=77K for (a) $q=0.005\times 10^{8}$ m$^{-1}$ and (b) $q=0.5\times 10^{8}$ m$^{-1}$.
The potential bias is U=5 meV and the Fermi energy $E_F=0$. The details at low frequency for small $q$ is shown in the inset of (a).}
\label{fig:fig1}
\end{figure}

In intrinsic BLG where no net carrier exists, i.e., $N=0$ and the Fermi energy $E_F=0$, intraband
scattering is only allowed at non-zero temperatures. In Fig.\,\ref{fig:fig1}, we have shown that the
real part ($\varepsilon_r$, solid curve) and imaginary part ($\varepsilon_i$, dotted) of the dielectric
function versus the energy in an intrinsic system with potential bias $U=5$ meV
at a finite temperature 77 K for (a) $q=0.005\times 10^{8}$ m$^{-1}$ and (b) $q=0.5\times 10^{8}$ m$^{-1}$.
In the small $q$ case, the intraband
scattering introduces a dip for $\varepsilon_r$ and a peak for $\varepsilon_i$ of low energy as illustrated
in the insets. Consequently, there exist two plasmon modes, one Landau damped and one almost undamped. The
depth and the width of this real part dip increase with the temperature indicating the increase of
intraband scattering strength and also the energy of the undamped plasmon mode.
At $\omega=\sqrt{U^2+q^4/16m^\ast}$, the threshold of interband
single-particle excitation continuum (SPEC), $\varepsilon_i$ steps up and a sharp peak of $\varepsilon_r$
is observed thanks to the flat bottom and top of the energy bands. This peak may introduce additional plasmon
modes and is similar to the case in gapped monolayer graphene \cite{wang}.
As $q$ increases, the $\varepsilon_r$ dip ($\varepsilon_i$ peak) due to intraband scattering shifts quickly to
the higher energy side while the $\varepsilon_r$ peak ($\varepsilon_i$ step) due to the interband scattering moves only slowly.
As a result, the well separated intra- and inter-band structures at small $q$ mix with each other and
then separate again when $q$ increase as shown in Fig.\,\ref{fig:fig1}(b).
%Fig. 1

%Fig. 2
\begin{figure}
\vspace*{-1cm}
\hspace*{-1cm}
\includegraphics*[height=150mm, width=95mm]{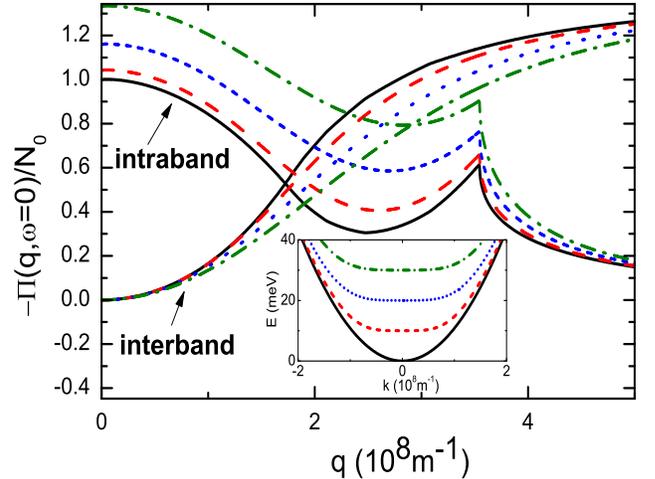}
\vspace{-8cm} \caption{The intra- and inter-band contributions to the real part of the
zero temperature propagator function $-\Pi$ in doped BLG with
carrier density $N=10^{12}$ m$^{-2}$ versus the wavevector $q$
are illustrated for $U=0$ (solid), 20 (dashed), 40
(dotted), and 60 (dash-dotted) meV.
The corresponding energy bands for different $U$ is shown in the inset.
$N_0=2m^\ast/\pi$ is the density of states of unbiased BLG.}
\label{fig:fig2}
\end{figure}

The effect of a bias on the zero-temperature propagator $-\Pi (q,\omega)$ in doped BLG \cite{hwan} with
a fixed carrier density $N=10^{12}$ m$^{-2}$ is studied in Fig.\,\ref{fig:fig2}. The interband contribution decreases
with the bias potential $U$ as the energy gap widens. The intraband contribution, on the contrary, increases with the bias
since the energy dispersion leads to an enhancement of the density of states near the Fermi energy. Here one of the characteristics in BLG
against in monolayer graphene \cite{wang2,hwan} is the strong back scattering of electrons on the Fermi surface
which results in an intraband peak at $q=2k_F=3.54\times 10^8$ m$^{-1}$. If the Fermi energy remains fixed as the bias increase,
the carrier density decreases and the intra- (inter-) band contribution at large (small) $q$ becomes less sensitive to the bias
and decreases (increases) with the bias.

%Fig. 3
\begin{figure}
\vspace*{-1cm} \hspace*{-0.5cm}
\includegraphics*[height=130mm, width=90mm]{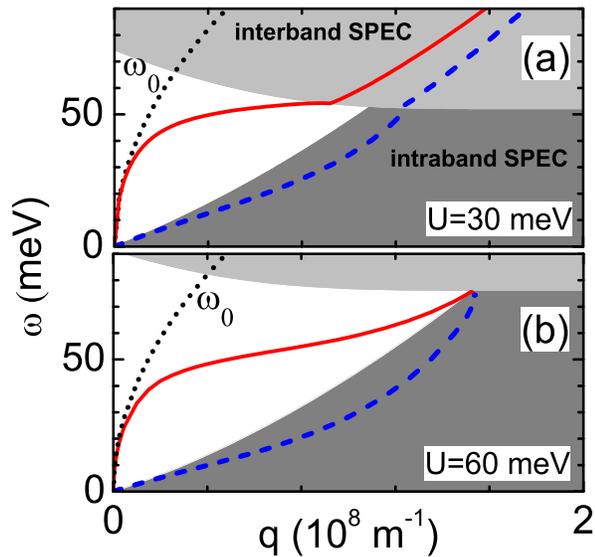}
\vspace{-5cm} \caption{The plasmon spectrum (solid curves for undamped or
slightly damped mode and dashed for Landau damped mode) and
single particle continuum spectrum
(light shadow for interband SPEC and dark shadow for intraband SPEC)
at zero temperature under potential bias (a) $U=30$ and (b) $U=60$ meV
are illustrated for BLG with $N=10^{12}$ m$^{-2}$.
The plasmon dispersion in the long wavelength limit,
$\omega_0=\sqrt{(e^2/4\pi\epsilon_0)/(N\pi q/m^\ast)}$ is also presented.}
\label{fig:fig3}
\end{figure}

\begin{figure}
\vspace*{-4.5cm} \hspace*{-0.7cm}
\includegraphics*[height=150mm, width=95mm]{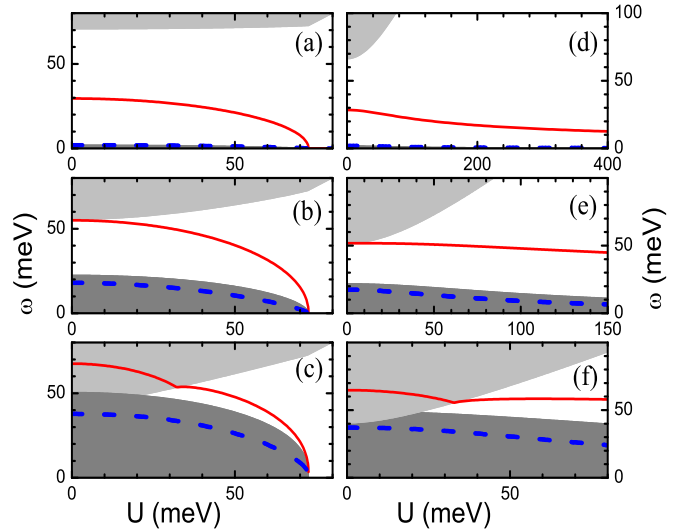}
\vspace{-3.5cm} \caption{The energy $\omega$ versus the potential bias $U$ of zero temperature plasmon modes
(solid for undamped and dashed for Landau damped) at fixed $E_F=36$ meV
(left panels) or at fixed $N=10^{12}$ m$^{-2}$ (right panels) for $q=0.05$ in (a) and (d), $0.5$ in (b) and (e),
and $1 \times 10^{8}$ m$^{-1}$ in (c) and (f), respectively.}
\label{fig:fig4}
\end{figure}

The plasmon modes are obtained by solving the zeros of the real part of the dielectric function
$\varepsilon_r(q,\omega)=0$ and the corresponding imaginary part $\varepsilon_i$ represents the damping
rate of the plasmon modes. In Fig.\,\ref{fig:fig3}, we plot the typical spectrum of plasmon modes (solid and dashed curves) in doped BLG at zero temperature under potential bias (a) $U=30$ meV and (b) $U=60$ meV.
The upper light shadow is the interband
SPEC edged at $\omega=\sqrt{U^2+k_F^4/m^{\ast 2}}/2+\sqrt{U^2+(k_F-q)^4/m^{\ast 2}}$
and the lower dark shadow is the intraband SPEC edged at
$\omega=\sqrt{U^2+(k_F+q)^4/m^{\ast 2}}-\sqrt{U^2+k_F^4/m^{\ast
2}}/2$. One undamped mode is located in the SPEC gap due to finite Fermi energy.
In the long wavelength limit, its dispersion is the same as that of Fermi 2D gas
of two valley, $\omega_0=\sqrt{(e^2/4\pi \epsilon_0)/(N\pi q/m^\ast)}$.
However, compared to the plasmon dispersion in BLG without bias
which is just slightly modified from that of Fermi 2D gas, this
dispersion is greatly softened for finite $q$ as also shown in Fig.\,\ref{fig:fig4}.
Our numerical analysis shows that this is a result of
the deformation near the bottom and top of the energy bands. Note that the lowered plasmon group velocity
may be helpful for making a stimulated plasmon oscillator \cite{rana}. A Landau damped mode is located just
below the intraband SPEC edge as usually happens in traditional 2D Fermi gas but is pushed to lower energy
at larger $q$. The undamped mode can enter into the interband SPEC and becomes a slightly damped mode in some cases
as shown in Fig.\,\ref{fig:fig3}(a) under $U=30$ meV or merges with the damped mode and disappears near the cross of
intra- and inter-band SPEC edges as shown in Fig.\,\ref{fig:fig3}(b) under $U=60$ meV.

%Fig. 4

With the plasmon spectrum in mind, we now explore how $U$ affects the energy
and damping properties of the modes.
In the left panels of Fig.\,\ref{fig:fig4}, we show $\omega$ versus $U$
at several typical $q$ when keeping the
Fermi energy constant. As in Fig.\,\ref{fig:fig3}, the light shadow indicates
the interband SPEC and the
dark shadow for the intraband one. At small $q$ as illustrated in (a),
there is one undamped plasmon mode
with energy located inside the SPEC gap of
which the width is about $2E_F$ and one damped mode of low
frequency. When $U$ reaches and passes $2E_F$, the Fermi level drops
below the conduction bottom and the
two plasmon modes merge and disappear. At larger $q$, the intraband SPEC edge shifts up
and the interband one shifts down for $U<2E_F$ and $\omega$ increases
as shown in (b) and also in
Fig.\,\ref{fig:fig3}. Then the two SPECs will merge and
the previous undamped plasmon mode enter the
interband SPEC and become slightly damped.
In this case, we may open the SPEC gap again by applying a
stronger bias and transfer the slightly damped plasmon mode
into a undamped as shown in (c). The $\omega$
versus $U$ curve forms a shoulder when it meets
the interband SPEC reflecting the strong coupling between
the single particle and collective excitations as also
shown in other cases \cite{wang,wang1,wang2}.

If $N$ remains constant as shown in the right panels,
the Fermi vector is also constant but the $E_F$
shifts up with $U$. This is clearly shown in (d)
by the interband SPEC edge of small $q$ which is
located near $\omega=2E_F$. The undamped plasmon mode
continues to exist as $U$ increases and its energy
varies slowly. This is because $E_F$ is always higher than
the conduction bottom with a constant
Fermi vector. $\omega$ decreases with $U$ as the effective mass near $E_F$ increases.
For a large $q$
the plasmon mode located inside the interband SPEC and
is slightly damped at small $U$, one can always make
it undamped by increasing the bias and widening the gap between
the intra- and inter-band SPECs as shown in (f).

When an external gate voltage is applied to a BLG, the carrier density varies
with the gate voltage as well as the energy gap. \cite{min,gava,mcca1,falk} Although $N$ and $U$ can be dependent on each other in a nontrivial way, our result suggests that
$\omega$ is proportional to $\sqrt{N}$ in almost the same way in both doped and undoped
BLG. This happens because $\omega$ is mainly determined by $N$ as illustrated in the
right panels of Fig.\,\ref{fig:fig4}. The variation of $U$ of small amount affects
$\omega$ only in a very limited scale. Nevertheless, as shown in Fig.\,\ref{fig:fig4}, the higher $U$ opens an wider energy gap in the
SPEC and prolongs the lifetime of the plasmon modes. In other words, a gate voltage can
vary the imaginary part of the dielectric constant at the plasmon energy and
the effect may be observed in experiments.

In summary, a potential bias can be applied between
the two graphene layers of a bilayer graphene with
the help of a gate voltage. We have studied the effect of
the potential bias on electronic correlations,
Coulomb screening, and collective excitations at both zero and finite temperature.
The potential bias
opens a gap in the single particle energy spectrum and makes
the semimetal bilayer graphene a semiconductor.
As a result the dielectric function for the Coulomb interaction and
the propagator function are modified
significantly. The potential bias also opens a gap in
the single-particle excitation spectrum and softens
the collective excitation modes. This may result in
undamped collective excitation modes that are observable
in experiments. In the single gate configuration, the doping and
gate voltage can vary the potential bias
and the carrier density of the bilayer graphene and manipulate
the energy and lifetime of the collective
excitation modes inside.

We acknowledge helpful discussions with D. S. L. Abergel. X. F. W.
acknowledges support from the startup fund for distinguished professors in Soochow University and T. C. acknowledges support from Canada Research Chair Program and the NSERC Discovery Grant.

\end{document}